\documentstyle[11pt, epsf]{article}
\newcommand{\bm}{\bibitem}

\begin{document}
\thispagestyle{empty}
\topmargin -1.0 cm
\headsep 1.5 cm
\textheight 22.0  cm
\normalbaselineskip = 24 true pt
\normalbaselines
\rightline{\large\bf SINP-TNP/96-07}

\rightline{\large\bf TIFR/TH/96-22}

\rightline{\large\bf April 1996}

\begin{center}
{\Large\bf
{ Matter induced $\rho-\omega$ mixing}}
\vskip 0.2 cm
{\sf Abhee K. Dutt-Mazumder \footnote{E-mail:
abhee@tnp.saha.ernet.in}, 
Binayak Dutta-Roy }\\
Saha Institute of Nuclear Physics, 1/AF  Bidhan Nagar\\
Calcutta - 700 064, India\\[1mm]
and\\[1mm]
{\sf Anirban Kundu \footnote{E-mail:akundu@theory.tifr.res.in}}\\
Tata Institute of Fundamental Research, Homi Bhabha Road,\\
Mumbai - 400 005, India.
\end{center}
\begin{abstract}
A novel mechanism for $\rho-\omega$ mixing induced by neutron-proton
asymmetry in nuclear matter is uncovered and the variation of the mixing angle
with the extent of asymmetry is presented.
%{\it{Keywords:}} $\rho^0$ meson, mixing, asymmetric\\
\end{abstract}

\newpage

The mixing of $\rho$ and $\omega$ mesons has attracted considerable
attention for quite some time. The phenomenon of the needle (the
narrow $\omega$) in the haystack (the broad $\rho$ resonance)
was one of the amusing consequences of this idea. This has also been invoked
in the context of the observed charge symmetry
violation in mirror nuclei --- commonly known as Okamato-Nolen-Schiffer 
anomaly \cite{oka,nolen}. Several authors have investigated the problem
and have attempted to explain the effect by constructing
the Charge Symmetry Violating (CSV) potential, taking 
$\rho^0-\omega$ mixing into account \cite{barr}. Controversies
in this field have sparked off interest in 
alternate sources of isospin symmetry violation. 
In ref. \cite{col}, 
$\rho-\omega$ mixing has been studied within the framework
of Vector Meson Dominance where the electromagnetic mixing
is considered, while the same has been investigated by Gao
{\em et al} \cite{gao} using an extended Nambu-Jona-Lasinio model.
Several authors also 
take an isospin violating vertex factor as a source
of the mixing \cite{gard} and others have implicated the u-d mass difference
in the quark loop as a contributor to the mixing \cite{cohen}.

However, the present treatment is somewhat different both in motivation 
and genesis from the earlier works, the main focus being the
study of the mixing arising from the
isospin symmetry breaking stemming from the nuclear medium through which
the hadron propagates.
Here significant mixing results, as we  shall show, when the 
$p\leftrightarrow n$ symmetry of the nuclear matter is broken. 
The interaction Lagrangian involving the nucleons and the 
vector mesons may be written as 
\begin{equation}
{\cal L}_{int} = g_{\rho} [{\bar{p}} \gamma _\mu p -
\bar n \gamma _\mu n] \rho^\mu_0 +
\sqrt{2}g_\rho[{\bar{p}}\gamma _\mu n \rho^{\mu +} + h.c ] +
g_\omega 
[{\bar{n}}\gamma_\mu n +{\bar{p}}\gamma_\mu p ] \omega ^\mu
\end{equation}                                                   
which is isospin conserving.
It may be noted that the coupling of the isovector $\rho ^0$ 
with the neutron and the proton has opposite sign in contrast to what 
obtains for the isoscalar $\omega$.
As a consequence, when we evaluate
the self-energy diagrams relevant for $\rho^0 -\omega $ mixing, 
the neutron and proton loops contribute with opposite signs.
As long as the ground state is symmetric under $p\leftrightarrow n$
interchange their contributions cancel exactly (as is the case for vacuum or for 
symmetric nuclear matter) as must be the case to the extent that 
isospin symmetry is respected not only by the Lagrangian but also
by the medium. On the other hand
if the ground state symmetry is broken {\em i.e.}
number of neutron and proton in nuclear matter 
differ, then the corresponding Fermi momenta, $k_F^p$ and  $k_F^n$, 
will be unequal whereby the cancellation does not take place and 
$\rho^0-\omega$ mixing occurs.

The form of the nucleon propagator in nuclear matter may be expressed as
\begin{equation}
G(k) = G_F(k)  + G_D(k)
\end{equation}
where
\begin{equation}
G_F(k) =
(k\!\!\!/ +M^\ast)[\frac{1}{k^2-M^{\ast2}+i\epsilon}]
\end{equation}
and
\begin{equation}
G_D(k) = (k\!\!\!/ + M^\ast) [\frac{i\pi}{E^{\ast}(k)}
\nonumber\\ \delta (k_0-E^\ast (k))\theta(k_F-|\vec k |)]
\end{equation}
with $E^\ast (\vert \vec k \vert) = \sqrt {\vert \vec k \vert^2 + M^{\ast 2}}$,
$M^\ast$ being the effective mass of the nucleon in nuclear matter \cite{se}.
This density dependent shift of the nucleon mass $(M\rightarrow M^\ast)$
in nuclear matter is estimated, for the present purpose, by solving the 
self-consistent equation within the framework of the Walecka Model 
\cite{se,wa}.
The first term in $G(k)$, namely $G_F(k)$, 
is the same as the free propagator of a spin $\frac{1}{2}$
fermion, except for the fact that the effective mass of the nucleon
in nuclear matter
is to be used, while the second part, $G_D(k)$, 
involving $\theta(k_F-|\vec k|)$,
arising from Pauli blocking, describes the modifications 
of the same in the nuclear matter at zero temperature \cite{se}, as it
deletes the on mass-shell propagation of the nucleon in nuclear
matter with momenta below the Fermi momentum. Finite temperature effects 
may be incorporated by replacing the Heaviside $\theta$-function by
the Fermi distribution, but the main features to be exposed in
this letter are not modified significantly.
In general, the vector-boson induced polarization of the medium,
to one-loop, may be written as
\begin{equation}
\Pi_{\mu\nu}^{lm}(q^2)=\frac{-i}{(2\pi)^4}
\int {d^4}k~{\rm Tr}[i\Gamma^l_\mu iG(k+q)
{i\Gamma^m_\nu} iG(k)]
\end{equation}
where $\Gamma$ represents the $\rho(\omega)\bar N N$
vertex factor and $l,m$ are the
isospin indices. Just like the nucleon
propagator, the polarization function can also be expressed as a sum
of two parts, 
\begin{equation}
\Pi_{\mu\nu}(q^2)= \Pi_{\mu\nu}^F(q^2) + \Pi_{\mu\nu}^D(q^2)
\end{equation}
where the free (Dirac) part $\Pi^F_{\mu\nu}$ comes from the product
$G^F(k+q)G^F(k)$, and the dense part $\Pi^D_{\mu\nu}$ comes from $[
G^F(k+q)G^D(k)+G^D(k+q)G^F(k)]$. The fourth combination,
$G^D(k+q)G^D(k)$, does not contribute in the region of
stability which concerns us here \cite{ch}. 
Formulating the suitable Dyson equations
for the dressed propagators, one arrives \cite{ch} at the condition 
for finding their poles:
\begin{equation}
\epsilon(q)=0
\end{equation}
where the dielectric function $\epsilon(q)$ is the 
product of transverse and longitudinal dielectric functions:
\begin{equation}
\epsilon(q)=\epsilon_L(q) \epsilon_T^2(q)
\end{equation}
and individual zeros for $\epsilon_L$ or $\epsilon_T$ gives the poles
for the longitudinal or the transverse modes respectively. In our notation,
taking $q=(q_0,|\vec q|,0,0)$, $(\mu,\nu)=(0,1)$ pertains to
the longitudinal and $(\mu,\nu)=(2,3)$ to the transverse mode.

For the unmixed case
\begin{equation}
\epsilon _{T(L)} = (1-\frac{\Pi^{T(L)}_{vv}}{q^2-m_v^2})
\end{equation}
where $v=(\rho,\omega)$ is the relevant vector meson. In the static limit ($|\vec q|
=0$), the effective mass of the vector meson is given by
\begin{equation}
q_0^2=m_v^2+\Pi_{vv}^T=m_v^2+\Pi_{vv}^L.
\end{equation}
where $\Pi^L_{vv}=-\Pi^{(v)}_{00}+\Pi^{(v)}_{11}$ and 
$\Pi_{vv}^T=\Pi^{(v)}_{22}=\Pi^{(v)}_{33}$.

When mixing takes place between $\rho$ and $\omega$, the polarization
tensor becomes a $8\times 8$ matrix, with nonzero $\Pi^{\rho\omega}$.
However, one can put it in a block diagonal form corresponding
to longitudinal and transverse sectors. Because of the isotropy 
of nuclear matter,
the transverse sector can further be split into  a direct sum of two 
$2\times 2$ matrices, with $\Pi_{22}=\Pi_{33}$. 
We are interested only in the static limit; therefore, it will
suffice to work with the simpler transverse part only, putting 
$|\vec q|=0$. In this case
the dielectric function is given by the determinant of
\begin{equation}
\pmatrix {1-{\Pi_{\rho\rho}^T\over q_0^2-m_{\rho}^2}&
{\Pi^T_{\rho\omega}\over q_0^2-m_{\rho}^2}\cr 
{\Pi^T_{\omega\rho}\over q_0^2-m_{\omega}^2}& 
1-{\Pi_{\omega\omega}^T\over q_0^2-m_{\omega}^2} }
\end{equation}
($\Pi_{\rho\omega}=\Pi_{\omega\rho}$), and the shifted masses 
are found from the zeros of $\epsilon_T$, {\em i.e.}
from the following equation:
\begin{equation}
(q_0^2 - m_\rho^2-\Pi^T_{\rho\rho})
(q_0^2 - m_\omega^2-\Pi^T_{\omega\omega})
-(\Pi^T_{\rho\omega}\Pi^T_{\omega\rho})=0
\end{equation}
One notes that the solutions of eqn. (12) must be identical with those
obtained for the longitudinal mode, since at the static limit there cannot
be any preferred direction. We will omit the superscript $T$ in our future
discussions, keeping in mind that at the static limit, $\Pi_{00}=0$ and
$\Pi_{11}=\Pi_{22}$. In eq. (12),
the left-hand side is nothing but the determinant of the matrix
$q_0^2{\bf 1}-{\cal M}$, where ${\cal M}$ is the one-loop modified mass
matrix
\begin{equation}
{\cal M}=\pmatrix {\Pi_{\rho\rho}+m_{\rho}^2 & \Pi_{\rho\omega}\cr
\Pi_{\omega\rho} & \Pi_{\omega\omega}+m_{\omega}^2}
\end{equation}

The mixing angle $(\theta)$  between $\rho^0$ and $\omega$ fields can be
obtained from 
\begin{equation}
\tan 2\theta={2\Pi_{\rho\omega}\over (m_{\omega}^2-m_{\rho}^2)
+(\Pi_{\omega\omega}-\Pi_{\rho\rho})}
\end{equation}

As has already been mentioned, the polarization tensor consists
of two terms, one coming from the dense part describing the polarization
of the Fermi sea, while the other arises out of the nucleon-antinucleon
excitation, {\em i.e} the effect of the Dirac Vacuum. The evaluation of 
$\Pi^D(q^2)$ is straightforward as there is a finite upper limit
involved in the loop integral \cite{abhee1}. 
However, the free part $\Pi^F(q^2)$,
needs to be regularized.
We use the method of dimensional regularization and employ the
following subtraction scheme \cite{abhee2}

\begin{equation}
\partial^n\Pi^F(q^2)/\partial(q^2) ^n\vert _{M^\ast\rightarrow M,
q^2=m_v^2}=0 (n=0,1,2...,\infty ). 
\end{equation}

It may be noted, 
from (1) that whereas the $\omega\bar p p$ and 
$\omega \bar n n$ vertex factors come with the same sign, the signs of the
$\rho^0 \bar n n$ and $\rho^0 \bar p p$ vertex factors differ. This ensures 
the complete cancellation of $\rho-\omega$ mixing amplitudes coming from
the proton and the neutron loops 
for symmetric nuclear matter ( as well as for vacuum).
However, $\Pi^{\rho\omega}_{\mu\nu}(q^2)$
acquires a non-zero value whenever the ground state is not symmetric under
$p\leftrightarrow n$ as has already been mentioned earlier. 
This comes solely from the loop integrals
relevant for the density-dependent part, as the integrals have
different upper limits, {\em viz}, $k_F^n$ and $k_F^p$, corresponding to
neutron and proton loops,  leading to a partial cancellation only.

We present our results in Figs. 1 and 2. We have taken $M_n=M_p=938$ MeV,
and normal nuclear matter density $(\rho_0 = 0.17~~fm^{-3})$. The
values of the coupling constants $g_\rho$ and $g_\omega$ are taken 
to be 2.71 and 10.1 respectively as estimated from Vector Meson Dominance
(VMD) fits \cite{sakurai,sh}.

In Fig.(1) we observe that in symmetric matter, mixing does not take place,
while with increasing asymmetry the mixing becomes appreciable. We present
here the results
pertaining to the role of mixing on the masses
at three times the normal nuclear matter density in 
order to make the effect stand out in clear relief.

It may be noted that the extent of mixing, as expected, varies appreciably
with asymmetry parameter $(\alpha=\frac{N_n-N_p}{N_p+N_n})$.
While the mass for $\rho^0$, taking mixing into account, increases 
with $\alpha$, that of $\omega$ exhibits an opposite trend as is
expected for the eigen values of a mass matrix. Also one should 
appreciate the fact that the mass splitting of $\rho-\omega$ in matter
is larger by an order of magnitude compared to what is observed
in vacuum. In addition an overall decrease of $\rho-\omega$ meson masses
are observed. As far as the mixing is concerned, the effect of asymmetry is 
more pronounced for the case of $\rho^0$ than for $\omega$. No mixing, however,
takes place in the symmetric limit for reasons explained in the text 
already as we are not including electromagnetic
or other such sources of mixing in order to focus attention 
on the mechanism at hand.
In normal nuclear density also such an effect on masses due to 
mixing is observed but on a lower scale.
However, even at normal nuclear matter density, we find a significant
value of the mixing angle $(\theta)$ which, starting from zero,
increases almost linearly with asymmetry parameter 
$(\alpha)$ as shown in Fig. 2.

The value of the mixing amplitude here
is ten times larger, 
even at normal nuclear matter density, 
than what one estimates for electro magnetic mixing\cite{col}. 
In matter, even though the mass difference 
of $\rho$ and $\omega$ meson (giving the energy denominator) 
increases by an order of magnitude the mixing
amplitude is also quite large, therefore we do  arrive at mixing angles
comparable to those obtained from the more conventional sources.

The qualitative and dramatic manifestation of the mixing of $\rho$ and 
$\omega$ shows up in the two pion decay channel. 
The mixing results in principle 
to a narrow spike arising from the $\rho$ component in the mixed 
$\omega$ (which has a small width). 
In nuclear matter, however, this mode is not useful as the pions, 
being strongly interacting, would rescatter and lose the initial information. 
However, the dilepton mode or the decay channel $\pi + \gamma $ could
possibly reveal the effect of mixing.

To conclude and summarize, we stress that the mixing of $\rho^0-\omega$ may
also arise from the isospin asymmetric nuclear matter which is completely
a matter induced effect arising from the two different upper limit
involved in the density dependent part of the polarization tensor
$(\Pi^D(q^2))$  {\em viz.} the
respective Fermi momenta $k_F^p$ and $k_F^n$ in the proton and neutron loop.
In the limit of 
complete asymmetry, and at densities available in the core of neutron stars
as well as in relativistic heavy ion collision, this shift can be nearly
50 Mev for $\rho^0$ and 25-Mev for $\omega$, from their unmixed values.
It may be noted that even in the absence of mixing, the mass
of $\omega$ and $\rho^0$ gets shifted in the medium. The 
mixing amplitude is an 
order of magnitude larger than that for electromagnetic mixing, even 
at normal nuclear matter density, which makes the phenomenon interesting
and worth investigating.

\newpage
\centerline{Figure captions}
%\centerline{\epsffile{mixfig1.ps}}
%\centerline{\epsffile{mixfig2.ps}}
%\centerline{Fig.1}
\begin{enumerate}
\item {\sf Fig. 1: Variation of vector meson effective masses against asymmetry 
parameter $(\alpha)$ at
three times normal nuclear matter density. Solid line and dotted line represent
masses with and without mixing respectively.}
\item {\sf Fig. 2: Variation of mixing angle (${\theta ^0}$) against 
asymmetry parameter at normal nuclear matter density.}
\end{enumerate}
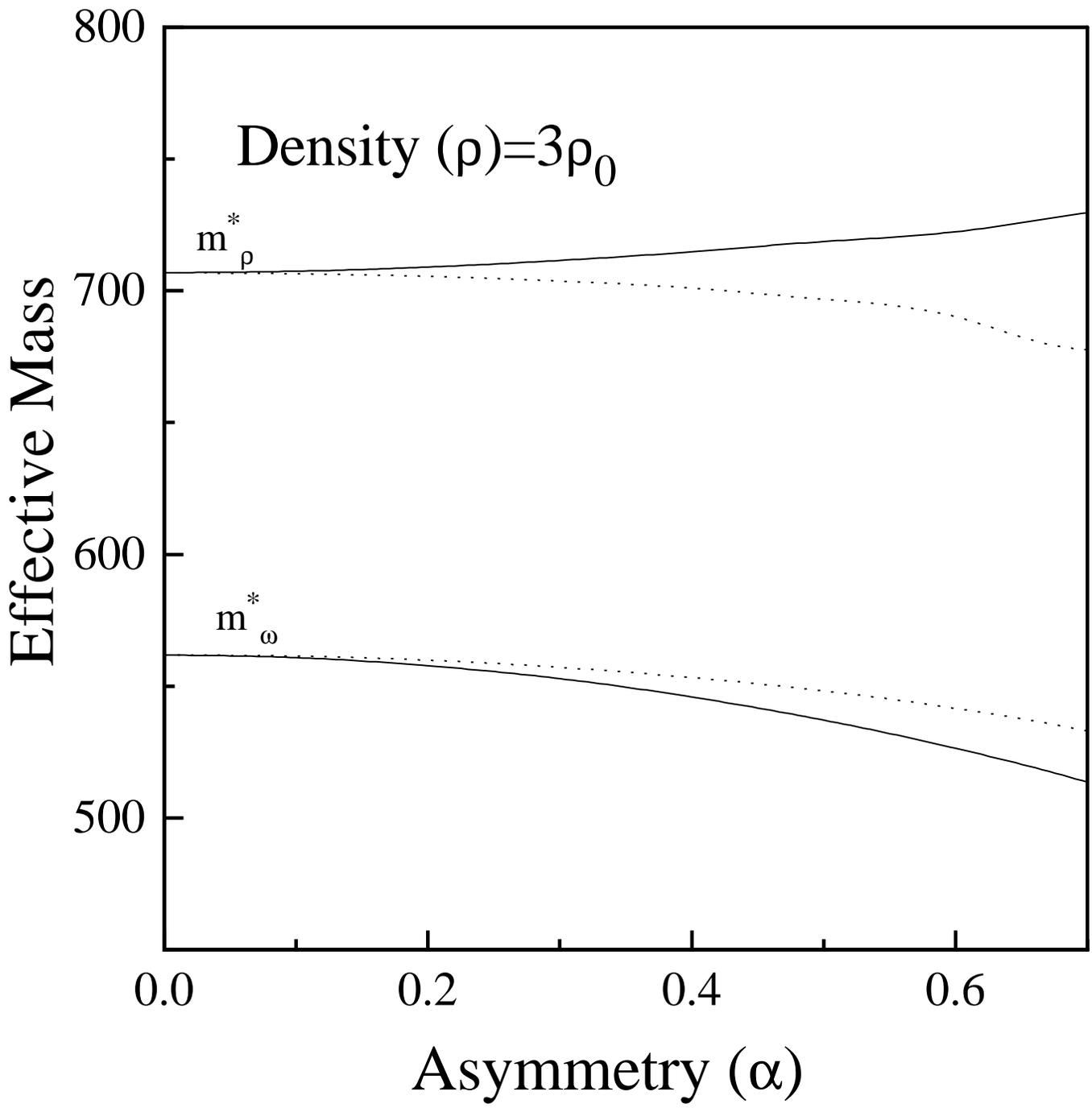
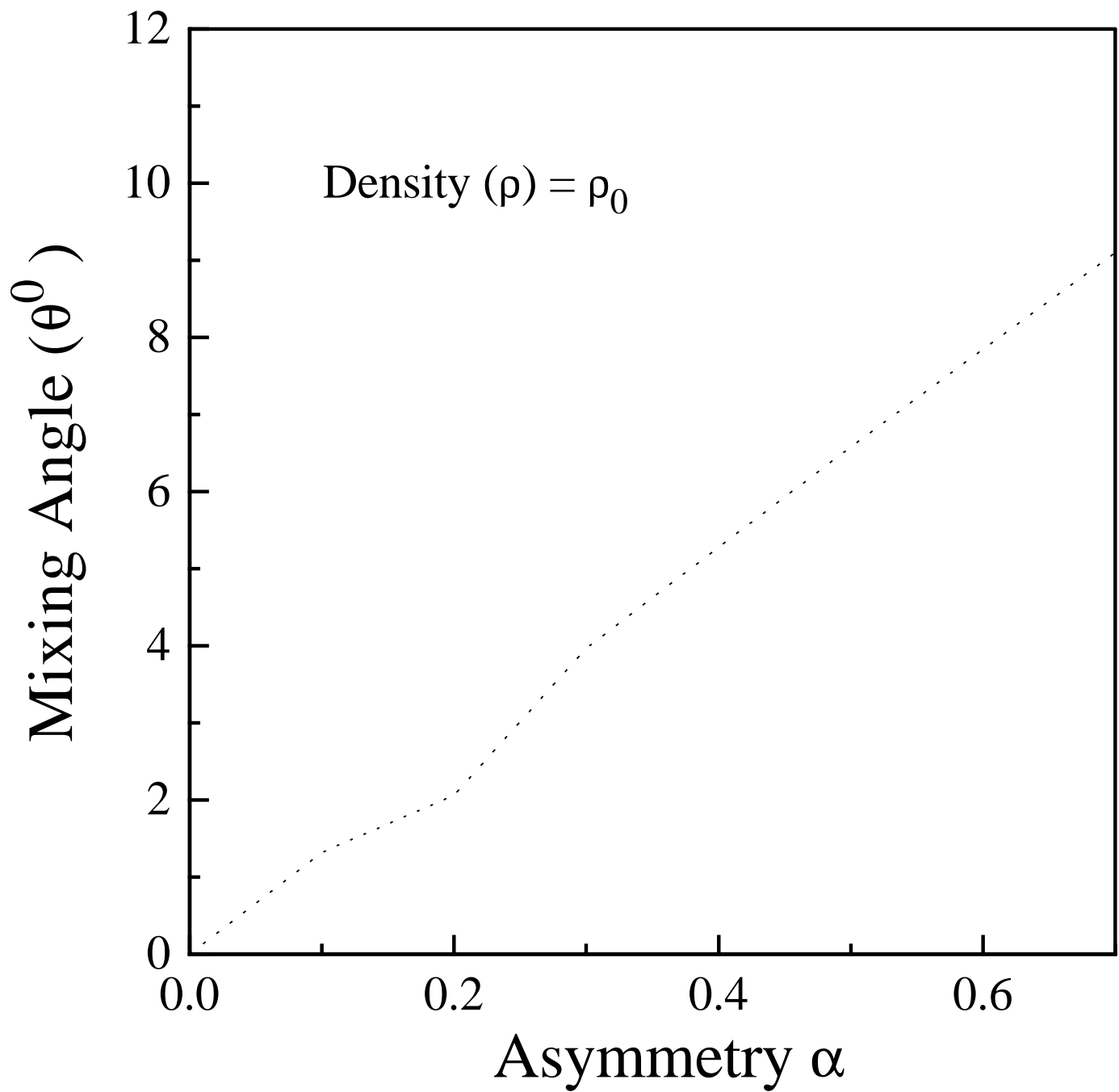
\end{document}